\documentclass{jpsj3}
\usepackage{txfonts}

\def\gsim{\mathop {\vtop {\ialign {##\crcr 
$\hfil \displaystyle {>}\hfil $\crcr \noalign {\kern1pt \nointerlineskip } 
$\,\sim$ \crcr \noalign {\kern1pt}}}}\limits}
\def\lsim{\mathop {\vtop {\ialign {##\crcr 
$\hfil \displaystyle {<}\hfil $\crcr \noalign {\kern1pt \nointerlineskip } 
$\,\,\sim$ \crcr \noalign {\kern1pt}}}}\limits}

\title{Robustness of Quantum Criticality of Valence Fluctuations}

\author{\name{Shinji \surname{Watanabe}}$^1$ and \name{Kazumasa \surname{Miyake}}$^2$}
\inst{$^1$Department of Basic Sciences, Kyushu Institute of Technology, Kitakyushu, Fukuoka 804-8550, Japan \\
$^2$Toyota Physical and Chemical Research Institute, Nagakute, Aichi 480-1192, Japan \\}

\abst{
The mechanism of emergence of robust quantum criticality in Yb- and Ce-based heavy electron systems 
under pressure is analyzed theoretically. 
By constructing a minimal model for quasicrystal Yb$_{15}$Al$_{34}$Au$_{51}$ and its approximant, 
we show that quantum critical points of the first-order valence transition of Yb appear 
in the ground-state phase diagram with their critical regimes being overlapped to be unified, giving rise to 
a wide quantum critical regime. 
This well explains the robust unconventional criticality observed in Yb$_{15}$Al$_{34}$Au$_{51}$ under pressure. 
We also discuss 
broader applicability of this mechanism to other Yb- and Ce-based systems 
such as $\beta$-YbAlB$_4$ showing unconventional quantum criticality.
}

\kword{robust quantum criticality, valence fluctuation, unconventional criticality, Yb- and Ce-based heavy electrons, quasicrystal}

\begin{document}
\maketitle

Quantum critical phenomena, which do not follow conventional spin fluctuation theory~\cite{Moriya,MT,Hertz,Millis}, have attracted much attention in condensed matter physics. 
Unconventional criticality commonly observed in paramagnetic metal phase in 
heavy-electron systems YbRh$_2$Si$_2$~\cite{Gegenwart} and $\beta$-YbAlB$_4$~\cite{Nakatsuji} 
challenges a paradigm of the magnetic quantum criticality (see Table~\ref{tb:QC}). 

Recently, it has been clarified theoretically by the present authors that critical valence fluctuation 
of Yb is the key origin of emergence of the new type of quantum criticality~\cite{WM2010}. 
We have found that  
almost dispersionless critical valence-fluctuation mode appears near ${\bf q}$=${\bf 0}$ in momentum space 
because of strong Coulomb repulsion of 4f holes at the Yb site. 
This gives rise to an extremely small characteristic temperature of critical valence fluctuations, $T_{0}$ 
with $T_{0}\ll T_{\rm K}$ where $T_{\rm K}$ is a characteristic temperature of heavy-electron systems, i.e., 
the so-called Kondo temperature. 
This makes experimentally-accessible low-temperature regime be located at ``high-temperature" $T/T_{0}\gg 1$ regime, which is the origin of emergence of anomalous criticality  
in physical quantities~\cite{WM2010}. 
Depending on the flatness of the critical valence fluctuation mode, the uniform magnetic susceptibility $\chi$ 
and the NMR/NQR relaxation rate $(T_{1}T)^{-1}$ shows 
$\chi\propto (T_{1}T)^{-1}\sim T^{-\zeta}$ with $0.5\lsim \zeta\lsim 0.7$. 
As shown in Table~\ref{tb:QC}, quantum valence criticality gives a unified explanation for 
the measured unconventional criticality.

\begin{table}[b]
\caption{New type of quantum criticality in uniform magnetic susceptibility $\chi$, 
NMR/NQR relaxation rate $(T_{1}T)^{-1}$, specific-heat coefficient $C/T$, and 
resistivity $\rho$. As for valence criticality, $T\gg T_0$ regime with $T_0$ being 
characteristic temperature of critical valence fluctuation is shown (see \cite{WM2010} for detail). }
\label{tb:QC}
\begin{center}
\begin{tabular}{lcccc}
\hline
\multicolumn{1}{c}{Material/Theory} & \multicolumn{1}{c}{$\chi$} & \multicolumn{1}{c}{$(T_{1}T)^{-1}$} & 
\multicolumn{1}{c}{$C/T$} & \multicolumn{1}{c}{$\rho$} \\
\hline
YbRh$_2$Si$_2$~\cite{Gegenwart} & $T^{-0.6}$ & $T^{-0.5}$ & $-\ln{T}$ & $T$ \\
$\beta$-YbAlB$_4$~\cite{Nakatsuji} & $T^{-0.5}$ & - & $-\ln{T}$ & $T^{1.5} \ \to \ T$  \\
Yb$_{15}$Al$_{34}$Au$_{51}$~\cite{Deguchi} & $T^{-0.51}$ & $\propto\chi$ & $-\ln{T}$ & $T$ \\
Valence criticality~\cite{WM2010} & $T^{-0.5\sim-0.7}$ &  $\propto\chi$ & $-\ln{T}$ & $T$ \\
\hline
\end{tabular}
\end{center}
\end{table}

Recently, newly synthesized heavy-electron metal Yb$_{15}$Al$_{34}$Au$_{51}$ 
with quasicrystal structure 
has been revealed to exhibit the common unconventional criticality~\cite{Deguchi,Watanuki}.
It has been reported that low-temperature behavior of $\chi$, $(T_{1}T)^{-1}$, 
specific-heat coefficient $C/T$, and resistivity $\rho$ is well explained by the 
theory of quantum valence criticality, as shown in Table~\ref{tb:QC}~\cite{Deguchi}.
This discovery suggests ubiquity of quantum valence criticality. 
The key origin is considered to be the locality of the critical valence fluctuation mode, 
which is basically ascribed to atomic origin at the Yb site not depending on the detail of 
lattice structures such as periodic lattice or quasicrystal.

In Table~\ref{tb:QC}, $\beta$-YbAlB$_4$ and Yb$_{15}$Al$_{34}$Au$_{51}$ show 
the quantum critical behavior without tuning control parameters such as 
pressure, magnetic field, or chemical doping. 
Here we have the following interesting question: Is it accidental or inevitable? 
To answer this question, we consider that a recent measurement may give a hint:  
Quantum criticality observed in Yb$_{15}$Al$_{34}$Au$_{51}$ 
is robust against hydrostatic pressure~\cite{Deguchi,Sato}.  
By applying pressure up to $P\sim 1.54$~GPa, critical behavior in physical quantities 
shown in Table~\ref{tb:QC} does not change.  
To get insight into the above fundamental question, here we try to understand the reason why 
robust criticality appears in Yb$_{15}$Al$_{34}$Au$_{51}$ under pressure.  

Thus the aim of this Letter is to clarify the key mechanism of the robustness of quantum criticality 
from the viewpoint of quantum valence criticality. 
Since the locality of valence fluctuation, i.e., charge transfer between the 4f electron at Yb 
and conduction electrons at surrounding atoms is considered to be important, we focus on the Yb-Al-Au cluster 
which is the basic unit of the quasicrystal. 
We show that quantum critical regime of valence fluctuations is extended to wide region in the phase diagram, 
which well explains robust criticality observed in Yb$_{15}$Al$_{34}$Au$_{51}$ under pressure. 

\begin{figure}
\includegraphics[width=6.0cm]{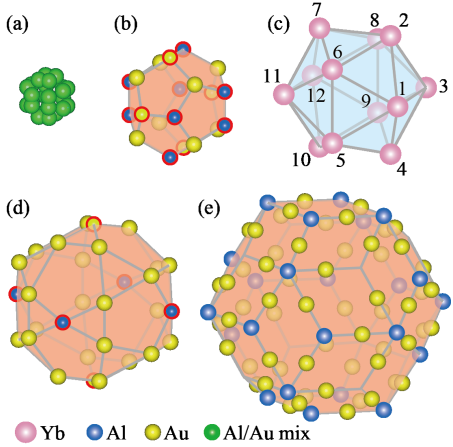}
\caption{(color online) Concentric shell structures of Tsai-type cluster in the Yb-Al-Au approximant:
(a) first shell, (b) second shell, (c) third shell, (d) fourth shell, and (e) fifth shell. 
The number in (c) indicates the $i$-th Yb site.
}
\label{fig:Yb_cluster}
\end{figure}

Let us start our discussion by analyzing the Yb-Al-Au cluster. 
Figure~\ref{fig:Yb_cluster} shows concentric shell structures of Tsai-type cluster 
in the Yb-Al-Au approximant, which is the basic structure of 
the quasicrystal Yb$_{15}$Al$_{34}$Au$_{51}$~\cite{Deguchi}. 
The approximant has periodic arrangement of the body-centered cubic structure whose unit cell 
contains the shell structures shown in Fig.~\ref{fig:Yb_cluster}(a)-(e). 
In the second shell, 12 sites are the Al/Au mixed sites 
(the sites framed in red in Fig.~\ref{fig:Yb_cluster}(b)) 
where the Al or Au atom exists  
with the rate of $62~\%/38~\%$, respectively~\cite{Ishimasa}. 
In the fourth shell, 6 sites are the Al/Au mixed sites 
(the sites framed in red in Fig.~\ref{fig:Yb_cluster}(d))  
where the Al or Au atom exists 
with the rate of $59~\%/41~\%$, respectively~\cite{Ishimasa}.
These rates are average values of the whole crystal. 
Hence the location of the Al and Au sites and the existence ratio can be 
different at the next-to-next concentric shells each other both in the quasicrystal and approximant. 
Thus the 2nd and 4th shells illustrated in Figs.~\ref{fig:Yb_cluster}(b) and 1(d), 
respectively, are such examples.

As noted above, 
almost dispersionless critical valence fluctuation mode is the key origin of emergence 
of the new type of quantum criticality shown in Table~\ref{tb:QC}. 
This implies that locality of valence fluctuation is essentially important. 
Namely, charge transfer between the Yb site and surrounding atoms 
is considered to play a key role, which is basically local. 
Hence we concentrate on the Yb-Al-Au cluster shown in Fig.~\ref{fig:Yb_cluster}.  

To construct a minimal model, 
we employ the result of recent experiment in Yb$_{15}$Al$_{34}$Au$_{51}$~\cite{Sato}: 
The same measurement as in Table~\ref{tb:QC} performed by replacing Al with Ga has revealed that 
quantum critical behavior in the physical quantities disappears. 
This suggests that conduction electrons at the Al site contribute to 
the quantum critical state. 
Hence we consider a simplest minimal model for the 4f-hole orbital at the Yb site and 
conduction-hole orbital at the Al site:  
\begin{eqnarray}
H&=&-\sum_{\alpha=2,5}\sum_{\langle \xi\nu\rangle\sigma}^{\rm (\alpha)}t_{\xi\nu}^{(\alpha)}\left(
c_{\xi\sigma}^{\dagger}c_{\nu\sigma}+{\rm h.c.}
\right)
-\sum_{\xi\sigma}^{(4)}\sum_{\eta}^{\rm (5)}t_{\xi\eta}\left(
c_{\xi\sigma}^{\dagger}c_{\eta\sigma}+{\rm h.c.}
\right)
\nonumber
\\
&+&\varepsilon_{\rm f}\sum_{i=1 \sigma}^{12}n_{i\sigma}^{\rm f}
+U\sum_{i=1}^{12}n_{i\uparrow}^{\rm f}n_{i\downarrow}^{\rm f}
+\sum_{i=1\sigma}^{12}\sum_{\eta}^{\rm (2,4,5)}V_{i\eta}
\left(
f_{i\sigma}^{\dagger}c_{\eta\sigma}+{\rm h.c.}
\right)
\nonumber
\\
&+&U_{\rm fc}\sum_{i=1\sigma}^{12}\sum_{\eta\sigma'}^{\rm (2,4)}
n_{i\sigma}^{\rm f}n_{\eta\sigma'}^{\rm c}
\label{eq:PAM}
\end{eqnarray}
where $f_{j\sigma}^{\dagger}$ $(f_{j\sigma})$ and $c_{j\sigma}^{\dagger}$ $(c_{j\sigma})$ 
are creation (anihilation) operators of the f hole and 
the conduction hole at the $j$-th site with spin $\sigma$, respectively, and 
$n^{\rm f}_{j\sigma}\equiv f_{j\sigma}^{\dagger}f_{j\sigma}$ 
and $n_{j\sigma}^{\rm c}\equiv c_{j\sigma}^{\dagger}c_{j\sigma}$. 
The first term represents the conduction-hole transfer on the 2nd and 5th shells, 
where 
$\sum_{\langle \xi\nu\rangle}^{(\alpha)}$ denotes the summation of the nearest-neighbor Al sites 
on the $\alpha$-th shell ($\alpha$=2: Fig~\ref{fig:Yb_cluster}(b) and $\alpha$=5: Fig.~\ref{fig:Yb_cluster}(e)). 
The second term represents the conduction-hole transfer between the 4th and 5th shells, 
where $\sum_{\xi\sigma}^{(4)}\sum_{\eta}^{(5)}$ denotes the summation of the nearest-neighbor Al sites on the 5th shell 
for each Al site on the 4th shell (Fig.~\ref{fig:Yb_cluster}(d)). 
The third and fourth terms represent the f-hole energy level $\varepsilon_{\rm f}$ and 
onsite Coulomb repulsion $U$ on the 3rd shell (Fig.~\ref{fig:Yb_cluster}(c)), respectively. 
The fifth term represents the hybridization $V_{i\eta}$ between f and conduction holes,  
where 
$\sum_{i=1\sigma}^{12}\sum_{\eta}^{(\alpha)}$ denotes the summation of the nearest-neighbor sites 
on the $\alpha$-th shell ($\alpha$=2, 4, or 5) for each $i$-th Yb site on the 3rd shell. 
 
The last term represents the inter-orbital Coulomb repulsion $U_{\rm fc}$.
This term has been shown theoretically to be essentially important to cause the quantum criticality of 
Yb-valence fluctuations~\cite{WM2010,M2007}. 
We note that in YbRh$_2$Si$_2$, Yb 3d-4f resonant photoemission measurement has revealed recently 
that Yb 5d electrons contribute to the energy band located at the Fermi level~\cite{Yasui}.
This strongly suggests importance of inter-orbital Coulomb repulsion between the 4f and 5d states due to its onsite nature. As for Yb$_{15}$Al$_{34}$Au$_{51}$, we expect that charge transfer between 4f and conduction states is also considerably influenced by the Yb 5d state which is considered not only to contribute to $U_{\rm fc}$ but also to hybridize with the conduction states at surrounding atoms since 5d wave function is spreading to a certain extent.  
Here, we introduced the $U_{\rm fc}$ term 
between the 3rd (Yb) shell and the conduction states at the surrounding Al sites  
to express this effect most simply as in eq.~(\ref{eq:PAM}) 
instead of introducing the 5d orbital at each Yb site explicitly. 
We note that quantum valence criticality in Table~\ref{tb:QC} is shown to appear 
on the basis of the periodic Anderson model with the $U_{\rm fc}$ term 
whose structure is essentially the same as eq.~(\ref{eq:PAM})~\cite{WM2010}.

Emergence of heavy-electron behavior in Yb$_{15}$Al$_{34}$Au$_{51}$ is ascribed to the strong Coulomb repulsion $U$ 
working on-site 4f holes at the Yb site. 
To determine the ground-state phase diagram of the model eq.~(\ref{eq:PAM}), 
here we employ the slave-boson mean-field approach 
in the limit of strong hole correlation, $U=\infty$~\cite{Read}, 
as a first step of analysis. 
To describe the state for $U=\infty$, 
we consider $Vf_{i\sigma}^{\dagger}b_{i}c_{\eta\sigma}$ instead of $Vf_{i\sigma}^{\dagger}c_{\eta\sigma}$ 
in eq.~(\ref{eq:PAM}) by introducing the slave-boson operator $b_{i}$ at the $i$-th site 
to describe the $f^0$ state and require the constraint 
$\sum_{i=1}^{12}\lambda_{i}(\sum_{\sigma}n_{i\sigma}^{\rm f}+b_{i}^{\dagger}b_{i}-1)$ 
with $\lambda_{i}$ being the Lagrange multiplier. 
We employ the mean-field treatment as $\overline{b_i}=\langle b_{i}\rangle$. 
For the $U_{\rm fc}$ term in eq.~(\ref{eq:PAM}), we employ the mean-field decoupling 
as $n_{i\sigma}^{\rm f}n_{\eta\sigma'}^{\rm c}\approx 
n_{i\sigma}^{\rm f}\langle n_{\eta\sigma'}^{\rm c}\rangle+\langle n_{i\sigma}^{\rm f}\rangle n_{\eta\sigma'}^{\rm c}
-\langle n_{i\sigma}^{\rm f}\rangle\langle n_{\eta\sigma'}^{\rm c}\rangle$. 
By optimizing the ground-state energy with respect to $\lambda_{i}$ and $\overline{b_i}$,  
$\partial\langle H\rangle/\partial\lambda_{i}=0$ and 
$\partial\langle H\rangle/\partial\overline{b_i}=0$, 
we obtain a set of the mean-field equations. 
\begin{eqnarray}
\sum_{\sigma}\langle f_{i\sigma}^{\dagger}f_{i\sigma}\rangle+\overline{b_i}^2&=&1,
\label{eq:MF1}
\\
\sum_{\eta\sigma}^{(2,4,5)}V_{i\eta}\langle f_{i\sigma}^{\dagger}c_{\eta\sigma}\rangle+\lambda_{i}\overline{b_i}&=&0,  
\label{eq:MF2}
\end{eqnarray}
for $i=1,.., 12$. 

Since we now consider the Yb-Al-Au cluster without periodic lattice structure, Fourier transformation 
to momentum space which diagonalizes the mean-field Hamiltonian 
is not available to solving eqs. (\ref{eq:MF1}) and (\ref{eq:MF2}). 
Here we calculate them by using the Slater matrix as follows. 

The ground state of the mean-field Hamiltonian obtained from eq.~(\ref{eq:PAM}) 
is given by 
$|\phi\rangle=|\phi_{\uparrow}\rangle \otimes|\phi_{\downarrow}\rangle$ 
with 
$
|\phi_{\sigma}\rangle=\prod_{k=1}^{N_{\sigma}}
\left(
\sum_{j=1}^{N}\phi_{jk}^{\sigma}a_{j\sigma}^{\dagger}
\right)
|0\rangle, 
$
where $N$ is the total number of sites and $N_{\sigma}$ the total number of holes with $\sigma$ spin 
in the system. 
Here, $\phi_{jk}^{\sigma}$ is the $N\times N_{\sigma}$ Slater matrix constituted of the $N_{\sigma}$ 
eigen vectors with the $N$ dimension, which corresponds to the eigen values of 
the $N\times N$ mean-field Hamiltonian matrix from the lowest one to the $N_{\sigma}$-th lowest one. 
The creation operator $a_{j\sigma}^{\dagger}$ is given by 
$f_{j\sigma}^{\dagger}$ $(c_{j\sigma}^{\dagger})$ at the $j$-th site 
in the 3rd shell (the 2nd, 4th, and 5th shell) in Fig.~\ref{fig:Yb_cluster}.  
The expectation value of product of $a_{j\sigma}^{\dagger}$ and $a_{l\sigma}$ is calculated as 
\begin{eqnarray}
G_{jl\sigma}=\frac{\langle\phi_{\sigma}|a_{j\sigma}^{\dagger}a_{l\sigma}|
\phi_{\sigma}\rangle}{\langle\phi_{\sigma}|\phi_{\sigma}\rangle}
=\sum_{j'=1}^{N}\sum_{l'=1}^{N}\phi_{jj'}^{\sigma}g_{j'l'}\phi_{l'l}^{\sigma}, 
\label{eq:G}
\end{eqnarray}
where the matrix $g$ is given by $g\equiv(^{t}\phi^{\sigma}\phi^{\sigma})^{-1}$. 
By using eq.~(\ref{eq:G}), $\langle f_{i\sigma}^{\dagger}f_{i\sigma}\rangle$ 
and $\langle f_{i\sigma}^{\dagger}c_{\eta\sigma}\rangle$ in eqs.~(\ref{eq:MF1}) and 
(\ref{eq:MF2}) are calculated. 
The calculation scheme is summarized as follows: 
1) First we assume a set of the mean fields $\lambda_{i}$ 
and $\overline{b_i}$ for $i=1,..,12$. 
2) Then the mean-field Hamiltonian is set. 
3) By diagonalizing the Hamiltonian matrix, we obtain the $N_{\sigma}$ eigen vectors corresponding 
to the eiven values from the lowest one to the $N_{\sigma}$-th lowest one, which constitute the Slater matrix 
$\phi^{\sigma}$. 
4) By using eq.~(\ref{eq:G}), we calculate totally 24 mean-field equations 
of eqs.~(\ref{eq:MF1}) and (\ref{eq:MF2}). 
5) By using the multi-variable Newton method or the iterative method, we obtain the set of 
mean-fields $\lambda_{i}$ and $\overline{b_i}$ for $i=1,..,12$ for the next step. 
The above procedure from 1) to 5) is repeated until the mean-fields which satisfy eqs.~(\ref{eq:MF1}) and 
(\ref{eq:MF2}) within the required accuracy are obtained. 

We consider the case that 
Al atoms are located at the Al/Au mixed sites in the 2nd shell and 4th shell 
with inversion symmetry with respect to the cluster center as shown 
in Figs.~\ref{fig:Yb_cluster}(b) and \ref{fig:Yb_cluster}(d), respectively. 
Due to this symmetry, the number of the mean fields can be reduced from 12 to 6 for $\lambda_{i}$ 
and $\overline{b_i}$, which makes calculation simpler. 

\begin{figure}
\includegraphics[width=8.5cm,bb=0 0 650 770]{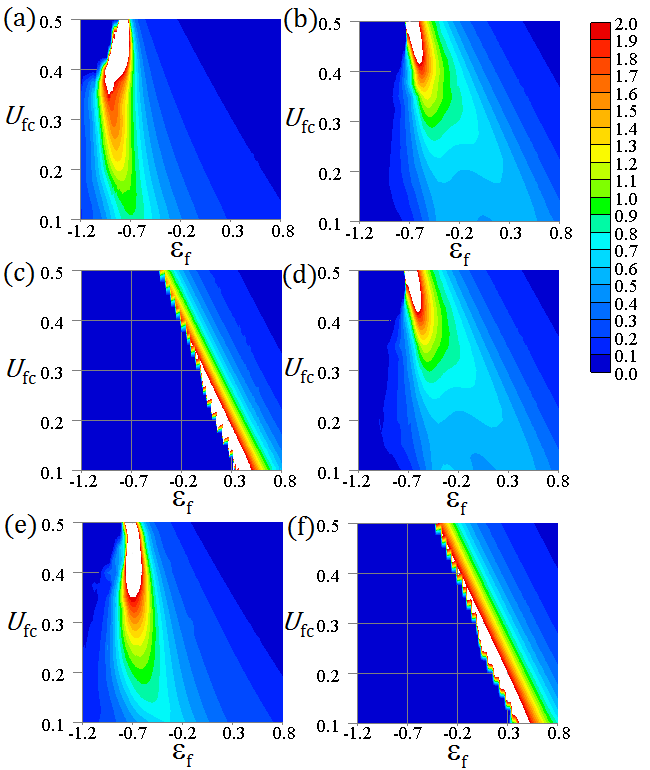}
\caption{(color online) Contour plot of valence susceptibility $\chi_{{\rm v}i}$ 
for (a) $i$=1, (b) 2, (c) 3, (d) 4, (e) 5, and (f) 6 
in the $\varepsilon_{\rm f}$-$U_{\rm fc}$ plane. 
Each white region shows  
diverging critical valence fluctuation arising from the VQCP for each $i$. 
}
\label{fig:chiv1_6}
\end{figure}

To understand the fundamental nature of this system, 
here we set $t_{\xi\nu}^{(2)}=t_{\xi\nu}^{(5)}=t_{\xi\eta}=t=1$, $V_{i\eta}=V=0.3$, and $U=\infty$ 
for a typical parameter set of heavy electron systems, 
as a first step of analysis. 
We consider the case that the total hole number   
$(N_{\uparrow},N_{\downarrow})=(24,24)$ in the Yb-Al-Au cluster with $N=54$ Yb and Al atoms in total, 
as shown in Fig.~\ref{fig:Yb_cluster}. 
By calculating the $\varepsilon_{\rm f}$ dependence of 
$\langle n_{i}^{\rm f}\rangle=\langle n_{i\uparrow}^{\rm f}\rangle +\langle n_{i\downarrow}^{\rm f}\rangle$ 
for each $U_{\rm fc}$ in the atomic configuration shown in Fig.~\ref{fig:Yb_cluster}, 
the valence susceptibility defined by 
$\chi_{{\rm v}i}\equiv -\partial\langle n_{i}^{\rm f}\rangle/\partial\varepsilon_{\rm f}$ 
is obtained, as shown in Fig.~\ref{fig:chiv1_6}.

This result can be understood qualitatively on the basis of the mean-field picture, as follows. 
Let us focus on the single f orbital at the $i$-th site on the 3rd shell (see Fig.~\ref{fig:Yb_cluster}(c)). 
When the f level is located at a deep position, i.e., 
$\varepsilon_{\rm f}$ is small enough in eq.~(\ref{eq:PAM}), 
one f hole is located at the $i$-th site with $\langle n^{\rm f}_{i}\rangle=1$, 
as a result of on-site strong hole correlations caused by $U=\infty$. 
As $U_{\rm fc}$ increases, at the point which satisfies 
\begin{eqnarray}
\varepsilon_{\rm f}+U_{\rm fc}\sum_{\eta}^{(2,4)}\langle n_{\eta}^{\rm c}\rangle\approx \mu, 
\label{eq:EfUfceq}
\end{eqnarray}
with $\mu$ being the chemical potential, $\langle n_{i}^{\rm f}\rangle$ shows a jump to 
the smaller $\langle n_{i}^{\rm f}\rangle$ value. 
Namely, first-order valence transition (FOVT) occurs 
since large $U_{\rm fc}$ forces holes pour into either the f level or the conduction orbital. 
As $\varepsilon_{\rm f}$ increases or $U_{\rm fc}$ decreases along the FOVT line, 
the value of the valence jump 
decreases and finally disappears at the quantum critical end point of the FOVT. 
This point is called 
the quantum critical point of the valence transition (VQCP),  
at which critical valence fluctuation diverges, i.e., $\chi_{{\rm v}i}=\infty$.
As further $\varepsilon_{\rm f}$ increases or $U_{\rm fc}$ decreases along the valence-crossover line 
extended from the FOVT line, the valence susceptibility $\chi_{{\rm v}i}$ is still enhanced, 
giving rise to the quantum critical regime in the $\varepsilon_{\rm f}$-$U_{\rm fc}$ plane.  
Intuitively, critical valence fluctuations are enhanced 
around the valence-crossover line near the VQCP because of the enhanced possibility whether for holes to stay 
at the f level or to move up to the Fermi level to avoid the energy loss due to $U_{\rm fc}$ 
(see eq.~(\ref{eq:EfUfceq})).  
 
When $V$ is set to be the larger value, the location of the VQCP is shifted to 
the smaller-$\varepsilon_{\rm f}$ and larger-$U_{\rm fc}$ direction 
in the $\varepsilon_{\rm f}$-$U_{\rm fc}$ phase diagram 
since larger $V$ promotes charge transfer between f level and conduction states.  
In the case of the smaller $V$, the VQCP is shifted to 
the larger-$\varepsilon_{\rm f}$ and smaller-$U_{\rm fc}$ direction. 

Now let us consider the 6 f-orbitals on the $i$-th site ($i$=1$\sim$6) on the 3rd shell 
(see Fig.~\ref{fig:Yb_cluster}(c)). 
Since Al is located with a certain rate at the Al/Au mixed sites on the 2nd and 4th shells, 
as shown in Figs.~\ref{fig:Yb_cluster}(b) and \ref{fig:Yb_cluster}(d), respectively, 
each f site is not equivalent.  
For instance, let us focus on the c-f hybridization between the 2nd and 3rd shells.
As seen in Figs.~\ref{fig:Yb_cluster}(b) and \ref{fig:Yb_cluster}(c), 
the f orbital at the $i$=1 Yb site hybridizes with 
the conduction orbitals at the three Al sites on the 2nd shell. 
On the other hand, the f orbital at the $i$=6 Yb site 
only hybridizes with the conduction orbital at one Al site on the 2nd shell. 
Other f orbitals at the $i$-th Yb site for $i=2,3,4$, and $5$ have the hybridization paths 
on the 2nd shell in between.
Namely, ``effective c-f hybridization strength" with the 2nd, 4th, and 5th shells 
is different each other for the $i$=1$\sim$6 Yb sites. 

As $\varepsilon_{\rm f}$ increases from a deep position, $\langle n_{1}^{\rm f}\rangle$ 
first changes at the FOVT as well as valence-crossover line (see Fig.~\ref{fig:chiv1_6}(a)) 
since the effective c-f hybridization is strongest among $i$=1$\sim$6. 
When $\varepsilon_{\rm f}$ exceeds the FOVT or valence-crossover line for $i=1$, 
charge transfer from the f orbital at the $i=1$ Yb site to the conduction orbitals 
at the surrounding Al sites occurs. 
Then the Fermi energy $\mu$ in eq.~(\ref{eq:EfUfceq}) is determined under the rearrangement of 
total holes in this system. 
As $\varepsilon_{\rm f}$ further increases, charge transfer occurs similarly 
when eq.~(\ref{eq:EfUfceq}) is satisfied for $\mu$ set under the rearranged total holes  
for $i=5,4,2$, and $3$, as shown in Figs.~\ref{fig:chiv1_6}(e), \ref{fig:chiv1_6}(d), \ref{fig:chiv1_6}(b), 
and \ref{fig:chiv1_6}(c), respectively. 
Finally, the f orbital at the $i=6$ Yb site which has the smallest effective c-f hybridization 
shows the FOVT with the VQCP, as shown in Fig.~\ref{fig:chiv1_6}(f). 

\begin{figure}
\includegraphics[width=7cm,bb=0 0 370 390]{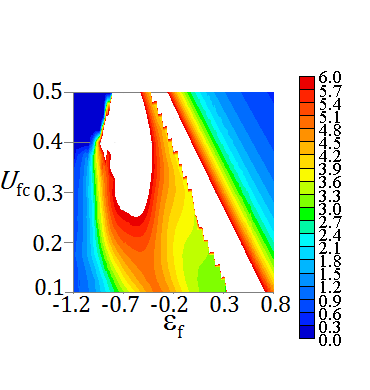}
\caption{(color online) Contour plot of total valence susceptibility $\chi_{\rm v}=\sum_{i=1}^{12}\chi_{{\rm v}i}$ 
in the $\varepsilon_{\rm f}$-$U_{\rm fc}$ plane.
White regions represent diverging critical valence fluctuations arising from the VQCP's 
for $i=1\sim 12$. 
Note that contour values are shown in larger scale than that in Fig.~\ref{fig:chiv1_6}. 
}
\label{fig:total_chiv}
\end{figure}

An important result is that VQCP's appear as islands in the $\varepsilon_{\rm f}$-$U_{\rm fc}$ phase diagram 
as shown in Fig.~\ref{fig:chiv1_6}, which makes critical regime enlarged.  
Actually, total valence susceptibility $\chi_{\rm v}\equiv\sum_{i=1}^{12}\chi_{{\rm v}i}$ 
shown in Fig.~\ref{fig:total_chiv} corresponding to experimental observation of criticality  
exhibit that critical valence fluctuations arising from each VQCP spot located closely are unified 
and hence the wide quantum critical regime appears in the $\varepsilon_{\rm f}$-$U_{\rm fc}$ plane. 
Note that although we now consider the case with inversion symmetry on the Al sites 
(see Fig.~\ref{fig:Yb_cluster}), in reality, absence of the symmetry gives rise to 12 VQCP spots 
but not 6 spots per an Yb-Al-Au cluster. 
This makes the critical regime be further enlarged in the $\varepsilon_{\rm f}$-$U_{\rm fc}$ phase diagram.

When pressure is applied to Yb-based materials, electrons located at the surrounding atoms approach the tail of 
wavefunction of the 4f electron at Yb, which makes the energy level of the crystalline electronic field 
increase~\cite{WM2011}. 
In the hole picture, this corresponds to 
decrease in $\varepsilon_{\rm f}$ in eq.~(\ref{eq:PAM}). 
Since $U_{\rm fc}$ is considered to increase under pressure in general, applying pressure to 
the Yb-based system corresponds to moving on the line toward the left-increasing direction 
in the $\varepsilon_{\rm f}$-$U_{\rm fc}$ phase diagram in Fig.~\ref{fig:total_chiv}.
Then in case that applying pressure follows the line in the enhanced critical regime, 
robust criticality is realized, which offers an explanation for robust criticality 
observed in Yb$_{15}$Al$_{34}$Au$_{51}$ under pressure~\cite{Deguchi}. 

Present analysis provides the core model both for quasicrystal and approximant: By further considering 
outer concentric shells in Fig.~\ref{fig:Yb_cluster}, quasicrystal structure is constructed, 
while by considering periodic arrangement of the concentric shells in Fig.~\ref{fig:Yb_cluster} 
as a unit cell, approximant is constructed. 
Although comparison of the electronic states in the bulk limit of both systems should be made 
for complete understanding of each system, the fundamental properties clarified here 
are considered to be unchanged even in the bulk limit 
since valence fluctuation is ascribed to atomic origin so that locality is essential. 
Namely, our result seems to be applied not only to systems with quasi-periodicity but also to those with periodicity of the lattice arrangement. 
Here we point out a possibility that difference in critical behavior 
between quasicrystal and approximant~\cite{Deguchi}  
may be ascribed to the location of the phase diagram: The former is located in the enhanced 
critical valence fluctuation regime as noted above and the latter seems slightly away from it. 

Emergence of a wide critical regime in the phase diagram offers a natural explanation for  
why quantum critical behavior was observed in materials without tuning control parameters in Table~\ref{tb:QC}.
As for $\beta$-YbAlB$_4$, four Yb atoms are located in the unit cell, 
which can be an origin of the robust criticality in this material,  
similarly to the above results. 
Here we note that rather short distance between Yb atoms along the $c$ axis 
$(\sim 3.4~\AA)$~\cite{Nakatsuji} is considered to contribute to the robust criticality. 
The Yb-Yb transfer via the conduction state at the B site promotes to 
widen the critical valence fluctuation regime in the phase diagram. 
We confirmed this tendency by the calculation in the periodic Anderson model with the $U_{\rm fc}$ term 
taking into account the effect of the f-f transfer. 
This can be understood intuitively as relaxation of the valence-fluctuating condition of eq.~(\ref{eq:EfUfceq})  
since $\varepsilon_{\rm f}$ has a certain width due to the f-f transfer effect.

To summarize, we have shown that robust criticality of valence fluctuation  
can appear in Yb-based heavy electron systems under ambient as well as applied pressure. 
This mechanism is considered to play a key role in Yb$_{15}$Al$_{34}$Au$_{51}$   
and is expected to have broader applicability also in the other Yb- and Ce-based heavy electron systems.


\begin{acknowledgment}


We thank N. K. Sato, T.~Ishimasa, K.~Deguchi, and T.~Watanuki for valuable discussions about experimental data.
One of us (S.W.) is supported by JASRI (Proposal 
Nos. 2012B0046 and 2013A0046). 
This work is supported by the Grant-in-Aid for Scientific Research (No. 24540378 and No.25400369)  
from the Japan Society for the Promotion of Science (JSPS).

\end{acknowledgment}


\begin{thebibliography}{9}
\bibitem{Moriya} T. Moriya: {\it Spin Fluctuations in Itinerant Electron Magnetism} 
(Springre-Verlag, Berlin, 1985). 
\bibitem{MT} T. Moriya and T. Takimoto: J. Phys. Soc. Jpn. {\bf 64} (1995) 960. 
\bibitem{Hertz} J. A. Hertz: Phys. Rev. B {\bf 14} (1976) 1165.
\bibitem{Millis} A. J. Millis: Phys. Rev. B {\bf 48} (1993) 7183. 
\bibitem{Gegenwart} P. Gegenwart, J. Custers, Y. Tokiwa, C. Geibel, and F. Steglich: 
Phys. Rev. Lett. {\bf 94} (2005) 076402, and references therein. 
\bibitem{Nakatsuji} Y. Matsumoto, S. Nakatsuji, K. Kuga, Y. Karaki, 
N. Horie, Y.~Shimura, T.~Sakakibara, A. H. Nevidomskyy, and P. Coleman: Science {\bf 331} (2011) 316,  
and references therein. 
\bibitem{WM2010} S. Watanabe and K. Miyake: Phys. Rev. Lett. {\bf 105} (2010) 186403.
\bibitem{Deguchi} K. Deguchi, S. Matsukawa, N. K. Sato, T. Hattori, K. Ishida, H. Takakura, and T. Ishimasa: Nature Mat. {\bf 11} (2012) 1013. 
\bibitem{Watanuki} T. Watanuki, S. Kashimoto, D. Kawana, T. Yamazaki, A. Machida, 
Y.~Tanaka, and T. J. Sato: Phys. Rev. B {\bf 86} (2012) 094201. 
\bibitem{Sato} N. K. Sato and K. Deguchi: private comunications.
\bibitem{Ishimasa} T. Ishimasa, Y. Tanaka, and S. Kashimoto: Phil. Mag. {\bf 91} (2011) 4218.  
\bibitem{M2007} K. Miyake: J. Phys.: Condens. Matter {\bf 19} (2007) 125201. 
\bibitem{Yasui} A. Yasui, Y. Saitoh, S.-I. Fujimori, I. Kawasaki, T. Okane, Y. Takeda, G.~Lapertot, 
G.~Knebel, T. D. Matsuda, Y. Haga, and H. Yamagami: Phys. Rev. B {\bf 87} (2013) 075131.
\bibitem{Read} N. Read and D. M. Newns: J. Phys. C: Solid State Phys. {\bf 16} (1983) 3273. 
\bibitem{WM2011} S. Watanabe and K. Miyake: J. Phys.: Condens. Matter {\bf 23} (2011) 094217.
\end{thebibliography}
\end{document}